\documentclass[11pt]{article}

\usepackage[a4paper,margin=1in]{geometry}
\usepackage[T1]{fontenc}
\usepackage[utf8]{inputenc}
\usepackage{amsmath,amssymb,amsthm}
\usepackage{bm}
\usepackage{graphicx}
\usepackage{hyperref}

\newcommand{\ii}{\mathrm{i}}
\newcommand{\dd}{\mathrm{d}}
\newcommand{\RR}{\mathbb{R}}

\newcommand{\orcid}[1]{\href{https://orcid.org/#1}{#1}}
\newtheorem{prop}{Proposition}
\newtheorem{remark}{Remark}

\title{Nonlocal Generalized Dirac Oscillators in (1 + 1) Dimensions}
\author{A.~Boumali\thanks{Corresponding author: \href{mailto:abdelmalek.boumali@univ-tebessa.dz}{abdelmalek.boumali@univ-tebessa.dz} (also: \href{mailto:boumali.abdelmalek@gmail.com}{boumali.abdelmalek@gmail.com}); ORCID: \orcid{0000-0003-2552-0427}.}\\
\small Laboratory of Theoretical and Applied Physics, Echahid Cheikh Larbi Tebessi University, Tebessa, Algeria}
\date{\today}

\begin{document}
\maketitle

\begin{abstract}
We propose a nonlocal extension of the generalized Dirac oscillator (GDO) in $(1+1)$ dimensions by replacing the
multiplicative interaction $f(x)$ with an integral operator $\hat F$ with kernel $f(x,x')$.
The resulting Dirac equation preserves an operator factorization and decouples into two nonlocal Schr\"odinger-type
(Sturm--Liouville) equations for the spinor components.
We derive explicit expressions for the associated supersymmetric partner kernels in terms of $f$ and its derivatives,
and we show that a complex-translation metric $\eta=e^{-\theta p_x}$ leads to a simple sufficient
\emph{kernel-level} pseudo-Hermiticity constraint,
$f(x+\ii\hbar\theta,x'+\ii\hbar\theta)=f^*(x',x)$, extending the familiar local complex-shift criteria.
To provide a transparent \emph{nonlocal-to-local} interpretation, we adapt the Coz--Arnold--MacKellar current-based
localization to each component equation, obtaining energy-dependent equivalent local potentials and multiplicative
Perey (damping) factors.
The mapping breaks down precisely at current zeros, thereby diagnosing the spurious solutions of the corresponding
nonlocal Schr\"odinger problem.
Finally, we illustrate the formalism with analytically tractable benchmarks (the local Dirac oscillator and a
translation-invariant kernel) and with a finite-rank separable model (Gaussian form factor) that reduces the
integro-differential problem to a small set of coupled ordinary differential equations and algebraic constraints.
\end{abstract}

% ==========================================================

\section{Introduction}
\label{sec:intro}

Nonlocality is a pervasive feature of effective single-particle descriptions obtained by eliminating (or averaging over)
unobserved degrees of freedom in an underlying many-body dynamics.
In coordinate space a nonlocal interaction acts as an integral operator,
\begin{equation}
  (\hat U\psi)(x)=\int_{\RR} \dd x'\,U(x,x')\,\psi(x'),
\end{equation}
so that the wave equation couples values of $\psi$ at distinct spatial points.
In nuclear reaction theory, nonlocality arises naturally from exchange/antisymmetrization effects and from channel
elimination; the Perey--Buck optical potential remains a paradigmatic phenomenological example \cite{PereyBuck1962}.

\par
A well-known manifestation of nonlocality is the \emph{Perey effect}: compared with the solution of a corresponding
local-equivalent equation, the wave function generated by an attractive nonlocal kernel is typically \emph{damped} in
the nuclear interior (and enhanced for repulsive kernels).  This behavior can often be summarized by a multiplicative
\emph{Perey factor} that relates the nonlocal wave function to the local-equivalent one and, in practice, modifies
interior-sensitive observables such as spectroscopic factors and transfer cross sections
\cite{Austern1965Perey,Rawitscher1985,FiedeldeyLipperheideRawitscherSofianos1992,TitusNunes2014}.
Recent analyses emphasize that the Perey factor is not universally accurate---its validity depends on the range and
energy dependence of the nonlocality and on the reaction mechanism---but it remains a useful diagnostic for gauging
the impact of localization approximations \cite{Albelleh2022NonlocalSources,TitusNunes2014}.

More generally, nonlocal kernels are routinely encountered in microscopic folding models and in dispersive optical
potentials, where they encode correlations beyond a static mean field and provide a bridge between reaction
observables and structure information \cite{MahauxSartor1991,Mahzoon2014PRL,Morillon2024GlobalNonlocalDOM,AtkinsonDickhoff2024NeutronSkins}.
Nonlocality also appears naturally in relativistic (Dirac-type) settings and in inverse spectral problems for
Dirac systems with integral kernels \cite{DebowskaNizhnik2019,HeribanTusek2024}.

A recurring theme in the nonlocal literature is the \emph{transition from nonlocal to local} representations.
Historically this transition has been pursued through (i) velocity-dependent or energy-dependent local
approximations \cite{FrahnLemmer1957}, (ii) explicit localization schemes such as the Brieva--Rook approximation and
its refinements \cite{BrievaRook1977I,BrievaRook1977II,HaiderRafiRookBhagwat2023}, and (iii) fixed-energy
phase-equivalent constructions \cite{BauhoffVonGerambPalla1983,LovellAmos2000}.
These approaches are complemented by exact or numerically defined \emph{local equivalents} that reproduce the
scattering solutions of a given nonlocal equation at a chosen energy and quantify the accompanying wave-function
renormalization (Perey effect) \cite{Austern1965Perey,FiedeldeyLipperheideRawitscherSofianos1992}.

A particularly transparent local-equivalent construction is based on current/Wronskian identities.
For definiteness, consider a one-dimensional nonlocal Schr\"odinger-type equation,
\begin{equation}
  -\hbar^2\,\psi''(x) + \int_{\RR} \dd x'\,V(x,x')\,\psi(x')=\epsilon\,\psi(x),\qquad \epsilon=\hbar^2 k^2.
  \label{eq:intro-nonlocal}
\end{equation}
One seeks an energy-dependent local potential $V^{\rm eq}(x;\epsilon)$ and a multiplicative factor $A(x;k)$ such that
\begin{align}
  &-\hbar^2\,\psi_L''(x) + V^{\rm eq}(x;\epsilon)\,\psi_L(x)=\epsilon\,\psi_L(x),
  \label{eq:intro-local-eq}\\
  &\psi_N(x)=A(x;k)\,\psi_L(x), \qquad \lim_{x\to\infty}A(x;k)=1.
  \label{eq:intro-perey}
\end{align}
The factor $A$ is the \emph{damping function} (or Perey factor): for attractive nonlocalities it typically suppresses
the interior amplitude of $\psi_N$ relative to its local-equivalent counterpart \cite{Austern1965Perey}.
Coz, Arnold, and MacKellar developed a mapping that yields closed-form expressions for both $V^{\rm eq}$ and $A$
whenever a Jost pair of nonlocal solutions remains linearly independent \cite{Coz1970}.
The same framework clarifies the appearance of \emph{spurious solutions}: when the incoming/outgoing solutions lose
linear independence, the normalized current vanishes, $A$ crosses zero, and the equivalent local potential can develop
singularities \cite{Coz1970,FiedeldeySofianos1983}.

Independently, non-Hermitian relativistic Hamiltonians may nevertheless possess real spectra when they are
$\eta$-pseudo-Hermitian, i.e.\ when $H^\dagger=\eta H\eta^{-1}$ for some Hermitian metric operator $\eta$
\cite{Mostafazadeh2002,Mostafazadeh2010}.
This is not merely a formal curiosity: when $\eta$ is positive definite, $H$ is similar to a Hermitian Hamiltonian
$h=\rho H\rho^{-1}$ with $\rho^\dagger\rho=\eta$.
Such similarity transformations frequently turn \emph{local} non-Hermitian interactions into \emph{nonlocal} Hermitian
ones, emphasizing that nonlocality and non-Hermiticity are, in many settings, two sides of the same effective
description.
In the context of generalized Dirac oscillators in $(1+1)$ dimensions, Dutta, Panella, and Roy used a complex
translation metric $\eta=e^{-\theta p_x}$ to construct broad families of complex interactions for which the Dirac
Hamiltonian is $\eta$-pseudo-Hermitian \cite{Dutta2013}.

\medskip
\noindent
The central objective of this work is to combine these two lines of development by formulating a
\emph{nonlocal generalized Dirac oscillator} (NLGDO) in $(1+1)$ dimensions and by providing a practical
nonlocal-to-local interpretation of its decoupled component equations.
Our main contributions are:
\begin{itemize}
\item[(i)] a sufficient \emph{kernel-level} pseudo-Hermiticity criterion for the nonlocal Dirac Hamiltonian under the
complex-translation metric $\eta=e^{-\theta p_x}$;
\item[(ii)] explicit formulas for the induced nonlocal supersymmetric partner kernels appearing in the second-order
component problems;
\item[(iii)] a component-wise local-equivalent mapping yielding energy-dependent local potentials and Perey-type
damping functions, together with a transparent breakdown criterion in terms of current zeros/spurious solutions.
\end{itemize}

The paper is organized as follows.
Section~\ref{sec:local} reviews the local GDO and the complex-translation metric.
Section~\ref{sec:nonlocal} defines the NLGDO, derives the decoupled nonlocal partner equations, and formulates a
sufficient kernel condition for $\eta$-pseudo-Hermiticity.
Section~\ref{sec:local-eq} adapts the Coz--Arnold--MacKellar local-equivalent construction to each component and
presents illustrative solvable and finite-rank examples.

% ==========================================================
\section{Background: local generalized Dirac oscillator and pseudo-Hermitian metrics}
\label{sec:local}

\subsection{Local generalized Dirac oscillator in $(1+1)$ dimensions}
The Dirac oscillator, introduced by Moshinsky and Szczepaniak, is obtained by a nonminimal substitution that produces
a linear (oscillator-like) coupling while preserving the first-order Dirac structure and exact solvability
\cite{MoshinskySzczepaniak1989,Sadurni2011DMOReview}.
A convenient one-dimensional generalization replaces the linear profile by an arbitrary (possibly complex) function
$f(x)$; this leads to the \emph{generalized Dirac oscillator} (GDO), which encompasses a broad class of solvable and
quasi-exactly solvable models and admits a natural supersymmetric factorization \cite{Sadurni2011DMOReview,Junker2020SUSYDirac}.

We work in the Pauli representation with
$\sigma_x=\begin{pmatrix}0&1\\1&0\end{pmatrix}$, $\beta=\sigma_z=\begin{pmatrix}1&0\\0&-1\end{pmatrix}$,
and momentum operator $p_x=-\ii\hbar\,\partial_x$.
The local GDO Hamiltonian reads
\begin{equation}
H_{\rm GDO}=c\,\sigma_x\Big(p_x-\ii\,\sigma_z f(x)\Big)+\beta\,m c^2
=\begin{pmatrix}
m c^2 & c p_x+\ii c f(x)\\
c p_x-\ii c f(x) & -m c^2
\end{pmatrix}.
\label{eq:HGDOlocal}
\end{equation}
For $\psi=(\psi_1,\psi_2)^T$, the Dirac equation $H_{\rm GDO}\psi=E\psi$ is equivalent to the coupled component system
\begin{equation}
\big(p_x+\ii f(x)\big)\psi_2=\frac{E-m c^2}{c}\,\psi_1,\qquad
\big(p_x-\ii f(x)\big)\psi_1=\frac{E+m c^2}{c}\,\psi_2.
\label{eq:components-local}
\end{equation}
Introducing the first-order operators
\begin{equation}
A=p_x-\ii f(x),\qquad A^\#=p_x+\ii f(x),
\end{equation}
one obtains the standard factorization/decoupling,
\begin{equation}
A^\#A\,\psi_1=\epsilon\,\psi_1,\qquad
AA^\#\,\psi_2=\epsilon\,\psi_2,\qquad
\epsilon=\frac{E^2-m^2 c^4}{c^2},
\label{eq:decouple-local}
\end{equation}
which reduces the spectral analysis to a pair of Schr\"odinger-type problems with supersymmetric partner potentials,
\begin{equation}
\left[-\hbar^2\frac{\dd^2}{\dd x^2}+V_\mp(x)\right]\psi_{1,2}=\epsilon\,\psi_{1,2},
\qquad
V_\pm(x)=f^2(x)\pm\hbar f'(x).
\label{eq:Vpm-local}
\end{equation}
In this representation $f(x)$ plays the role of a (possibly complex) ``superpotential'' and $(V_-,V_+)$ form a SUSY
pair.  The correspondence is useful both conceptually (spectral pairing, index theorems) and computationally (shape
invariance and algebraic construction of bound states) \cite{Sadurni2011DMOReview,Junker2020SUSYDirac}.
Throughout, the relativistic energy $E$ is recovered from the auxiliary eigenvalue $\epsilon$ via
$E=\pm\sqrt{m^2c^4+c^2\epsilon}$, with the sign selecting the positive/negative-energy branches.

\subsection{Pseudo-Hermiticity and complex-translation metrics}
A Hamiltonian $H$ is $\eta$-pseudo-Hermitian if $H^\dagger=\eta H\eta^{-1}$ with $\eta=\eta^\dagger$.
When $\eta$ can be chosen positive definite, $H$ is (quasi-)Hermitian and is similar to a Hermitian Hamiltonian
$h=\rho H\rho^{-1}$ with $\rho^\dagger\rho=\eta$ \cite{Mostafazadeh2002,Mostafazadeh2010}.
Pseudo-Hermiticity therefore provides a unifying framework that includes many PT-symmetric constructions as special
cases.

Motivated by complex-translation metrics, we focus on
\begin{equation}
\eta=e^{-\theta p_x},\qquad \theta\in\RR,
\label{eq:eta}
\end{equation}
which acts on sufficiently regular wave functions as a complex shift,
$(\eta\phi)(x)=\phi(x+\ii\hbar\theta)$.
For analytic interactions this choice turns pseudo-Hermiticity into a simple functional constraint.
This is the mechanism used in Ref.~\cite{Dutta2013} to generate families of complex generalized Dirac oscillators that
remain isospectral to their Hermitian counterparts.

\begin{prop}[Local complex-shift benchmark]
If $f(x)$ satisfies $f(x+\ii\hbar\theta)=f^*(x)$, then $H_{\rm GDO}$ in \eqref{eq:HGDOlocal} is
$\eta$-pseudo-Hermitian with $\eta=e^{-\theta p_x}$.
\end{prop}

\begin{remark}
Even when the local Hamiltonian is rendered Hermitian by a similarity transformation, the transformed operator may
acquire nonlocal (integral) terms if the metric $\eta$ is itself nonlocal.
This observation motivates treating nonlocality and pseudo-Hermiticity on the same footing in the generalized Dirac
oscillator setting.
\end{remark}

% ==========================================================
\section{Nonlocal generalized Dirac oscillator: definition, decoupling, and kernel pseudo-Hermiticity}
\label{sec:nonlocal}

\subsection{Definition}
We introduce nonlocality by replacing the multiplication operator $f(x)$ with an integral operator $\hat F$ acting on
scalar wave functions as
\begin{equation}
(\hat F\phi)(x)=\int_{\RR} \dd x'\, f(x,x')\,\phi(x').
\label{eq:Fkernel}
\end{equation}
The kernel $f(x,x')$ is allowed to be complex.
When $f(x,x')$ satisfies the Hermiticity condition $f(x,x')=f^*(x',x)$, the operator $\hat F$ is Hermitian; in that
case $H_{\rm NLGDO}$ below is a conventional (Hermitian) Dirac Hamiltonian with nonlocal interaction.
Our emphasis, however, is on pseudo-Hermitian settings in which nonlocality and complex structure coexist in a
controlled way.

The nonlocal generalized Dirac oscillator (NLGDO) Hamiltonian is defined by the formal replacement
$f(x)\mapsto \hat F$ in \eqref{eq:HGDOlocal},
\begin{equation}
H_{\rm NLGDO}=c\,\sigma_x\Big(p_x-\ii\,\sigma_z\hat F\Big)+\beta\,m c^2
=\begin{pmatrix}
 m c^2 & c p_x+\ii c\hat F\\
 c p_x-\ii c\hat F & -m c^2
\end{pmatrix}.
\label{eq:HNLGDO}
\end{equation}
The corresponding component equations generalize \eqref{eq:components-local} to
\begin{equation}
\big(p_x+\ii\hat F\big)\psi_2=\frac{E-m c^2}{c}\,\psi_1,\qquad
\big(p_x-\ii\hat F\big)\psi_1=\frac{E+m c^2}{c}\,\psi_2.
\label{eq:components-nonlocal}
\end{equation}

Spectral and inverse problems for Dirac-type operators with nonlocal potentials have also been investigated in
mathematically rigorous settings; see, e.g., Ref.~\cite{DebowskaNizhnik2019}.

\begin{remark}[Regularity and analyticity assumptions]
Throughout, we assume wave functions lie in a dense subspace of $L^2(\RR)$ (e.g.\ Schwartz space) on which $p_x$ is
defined, and that the kernel $f(x,x')$ is sufficiently smooth and decaying so that:
(i) $\hat F$ maps this domain into itself;
(ii) integration by parts in $x'$ is permitted with vanishing boundary terms; and
(iii) for complex shifts $x\mapsto x+\ii\hbar\theta$, $f$ admits analytic continuation in a strip of width at least
$\hbar|\theta|$ in both arguments and the relevant contour deformations cross no singularities.
These assumptions are standard in nonlocal scattering and in pseudo-Hermitian complex-shift constructions, and they are
made explicit here to justify the kernel manipulations used below.
\end{remark}

\subsection{Decoupling and explicit partner kernels}
A notable feature of the construction is that the first-order Dirac system \eqref{eq:components-nonlocal} retains a
factorization structure.
Define the nonlocal ladder operators
\begin{equation}
A=p_x-\ii\hat F,\qquad A^\#=p_x+\ii\hat F.
\label{eq:AAnl}
\end{equation}
Then
\begin{equation}
A^\#A=p_x^2+\hat F^2+\ii[\hat F,p_x],\qquad
AA^\#=p_x^2+\hat F^2-\ii[\hat F,p_x],
\label{eq:partner-operators}
\end{equation}
so the decoupled second-order equations become
\begin{equation}
A^\#A\,\psi_1=\epsilon\,\psi_1,\qquad
AA^\#\,\psi_2=\epsilon\,\psi_2,\qquad
\epsilon=\frac{E^2-m^2 c^4}{c^2}.
\label{eq:decouple-nonlocal}
\end{equation}
Writing \eqref{eq:decouple-nonlocal} in coordinate space yields a pair of integro-differential Schr\"odinger-type
problems,
\begin{equation}
-\hbar^2\,\psi_j''(x)+\int_{\RR} \dd x'\,V_j(x,x')\,\psi_j(x')=\epsilon\,\psi_j(x),
\qquad j=1,2,
\label{eq:nonlocal-schro}
\end{equation}
whose kernels $V_{1,2}$ play the role of nonlocal supersymmetric partners.
Under the assumptions of the previous remark (details in Appendix~A.1), one finds the explicit representation
\begin{equation}
V_{1,2}(x,x')=(f\star f)(x,x')\mp \hbar\big(\partial_x+\partial_{x'}\big)f(x,x'),
\qquad (f\star f)(x,x'):=\int_{\RR}\dd y\, f(x,y)f(y,x').
\label{eq:V12kernels}
\end{equation}
In the local limit $f(x,x')=f(x)\,\delta(x-x')$, Eq.~\eqref{eq:V12kernels} reduces to the standard partner potentials
$V_\pm(x)=f^2(x)\pm\hbar f'(x)$ in \eqref{eq:Vpm-local}.

\subsection{$\eta$-pseudo-Hermiticity for nonlocal interactions (kernel constraint)}
\label{sec:pseudo}
The Hermitian adjoint of $\hat F$ has kernel $f^\dagger(x,x')=f^*(x',x)$,
\begin{equation}
(\hat F^\dagger\phi)(x)=\int_{\RR} \dd x'\, f^*(x',x)\,\phi(x').
\end{equation}
Conjugation by $\eta=e^{-\theta p_x}$ shifts both arguments of the kernel:
\begin{equation}
(\eta\hat F\eta^{-1}\phi)(x)=\int_{\RR} \dd x'\, f(x+\ii\hbar\theta,x'+\ii\hbar\theta)\,\phi(x').
\label{eq:etakernel}
\end{equation}
Therefore, a sufficient condition for $\eta$-pseudo-Hermiticity of $H_{\rm NLGDO}$ is the \emph{kernel-level}
constraint
\begin{equation}
  f(x+\ii\hbar\theta,x'+\ii\hbar\theta)=f^*(x',x).
\label{eq:kernelcondition}
\end{equation}
Appendix~A.2 provides a detailed derivation.
In the local case, Eq.~\eqref{eq:kernelcondition} reduces to the complex-shift condition
$f(x+\ii\hbar\theta)=f^*(x)$ \cite{Dutta2013}.

\begin{prop}[Kernel criterion for pseudo-Hermiticity]
If the nonlocal interaction kernel satisfies \eqref{eq:kernelcondition} for some real $\theta$, then
$H_{\rm NLGDO}$ in \eqref{eq:HNLGDO} obeys $H_{\rm NLGDO}^\dagger=\eta H_{\rm NLGDO}\eta^{-1}$ with
$\eta=e^{-\theta p_x}$.
\end{prop}

\begin{remark}
Condition \eqref{eq:kernelcondition} interpolates between two familiar limits:
for $\theta=0$ it reduces to ordinary Hermiticity of the kernel, while for kernels concentrated near the diagonal
$x=x'$ it approaches the local complex-shift constraint of Ref.~\cite{Dutta2013}.
In this sense, \eqref{eq:kernelcondition} provides a natural pseudo-Hermitian extension of nonlocal interactions within
the generalized Dirac oscillator framework.
\end{remark}

% ==========================================================
\section{Local-equivalent formulation and illustrative models}
\label{sec:local-eq}

The decoupling in Section~\ref{sec:nonlocal} reduces the NLGDO problem to two nonlocal Schr\"odinger-type component
equations \eqref{eq:nonlocal-schro}.
To interpret these equations in familiar terms, we now construct \emph{local equivalents} for each component, following
the current-based strategy of Coz, Arnold, and MacKellar \cite{Coz1970}.
The resulting mapping produces, at a fixed energy, an energy-dependent local potential together with a multiplicative
Perey (damping) factor that quantifies the nonlocal modification of the wave function.

For clarity we present the mapping on the half-line $x\ge 0$; full-line scattering can be handled by treating the
left/right Jost systems separately \cite{Coz1970}.
Additional background on local-equivalent and phase-equivalent constructions can be found in
Refs.~\cite{PereyBuck1962,FiedeldeySofianos1983,Ross2015NonlocalTransfer,Bhoi2022PhaseEquivalent}.

\subsection{Nonlocal-to-local transition: localization, phase-equivalent potentials, and Perey factors}
Before turning to the explicit Coz mapping, it is useful to place the construction in the broader context of
nonlocal-to-local transitions.
A common starting point is that many physically motivated nonlocal kernels are short-ranged around the diagonal
$x\approx x'$, so that a gradient expansion yields local (but typically energy-dependent) operators.
Classic examples include the velocity-dependent Frahn--Lemmer interaction \cite{FrahnLemmer1957} and the
Brieva--Rook localization for exchange nonlocality in microscopic nucleon--nucleus potentials
\cite{BrievaRook1977I,BrievaRook1977II}, with modern reassessments and refinements in
Ref.~\cite{HaiderRafiRookBhagwat2023}.
At fixed energy, one may instead construct \emph{phase-equivalent} local potentials that reproduce the scattering phase
shifts of an inherently nonlocal interaction; see, e.g., the detailed transition analysis in
Ref.~\cite{BauhoffVonGerambPalla1983} and the inversion-based study of local/nonlocal equivalence in
Ref.~\cite{LovellAmos2000}.
In all of these approaches, the appearance of a multiplicative renormalization of the wave function---often referred to
as the Perey effect---is a robust diagnostic of nonlocality \cite{Austern1965Perey}.

The Coz--Arnold--MacKellar construction provides an \emph{exact} local equivalent (where defined) by exploiting the
Wronskian/current structure of a Jost pair of solutions.
It can be viewed as a constructive realization of the same idea that underlies Wronskian/inversion equivalent local
potentials discussed in, e.g., Ref.~\cite{FiedeldeyLipperheideRawitscherSofianos1992}: once two independent solutions of
the nonlocal equation are available, both the damping function and the equivalent local potential can be written
explicitly.

\subsection{Nonlocal component equation in Coz form}
Set $\epsilon=\hbar^2 k^2$ and define $U_j(x,x'):=V_j(x,x')/\hbar^2$.
Then \eqref{eq:nonlocal-schro} becomes
\begin{equation}
\left(\frac{\dd^2}{\dd x^2}+k^2\right)\psi_{N,j}(k,x)=\int_{0}^{\infty} \dd x'\,U_j(x,x')\,\psi_{N,j}(k,x'),
\qquad j=1,2.
\label{eq:nonlocal-coz-form}
\end{equation}
The corresponding local-equivalent equation is
\begin{equation}
\left(\frac{\dd^2}{\dd x^2}+k^2-U^{\rm eq}_j(x;k)\right)\psi_{L,j}(k,x)=0,
\qquad V^{\rm eq}_j(x;\epsilon)=\hbar^2 U^{\rm eq}_j(x;k).
\label{eq:local-eq}
\end{equation}
We emphasize that $U^{\rm eq}_j$ is, in general, energy dependent even when the original kernel $U_j$ is not; this is
the price paid for replacing an integral operator by a multiplicative potential at fixed energy.

\subsection{Jost-type solutions, normalized current, and damping function}
Let $\psi_{N,j,\pm}(k,x)$ denote a pair of independent nonlocal solutions normalized by outgoing/incoming
asymptotics as $x\to\infty$ (Jost-type solutions).
Assume a multiplicative map
\begin{equation}
\psi_{N,j,\pm}(k,x)=A_j(k,x)\,\psi_{L,j,\pm}(k,x),
\qquad \lim_{x\to\infty}A_j(k,x)=1,
\label{eq:transform}
\end{equation}
where $A_j$ is the damping function.
In scattering language, $A_j$ measures the local renormalization required to reconcile the interior behavior of the
nonlocal solution with a local equation that shares the same asymptotics.

Introduce the normalized Wronskian/current (Coz current)
\begin{equation}
J_{N,j}(k,x)=-\frac{1}{2\ii k}\Big[\psi_{N,j,+}(k,x)\,\partial_x\psi_{N,j,-}(k,x)
-\psi_{N,j,-}(k,x)\,\partial_x\psi_{N,j,+}(k,x)\Big].
\label{eq:JNdef}
\end{equation}
For the local equation \eqref{eq:local-eq} one has the corresponding normalized current $J_{L,j}\equiv 1$ for the
same asymptotic normalization.
Equation~\eqref{eq:transform} therefore implies the key identity
\begin{equation}
J_{N,j}(k,x)=A_j^2(k,x).
\label{eq:Aj2}
\end{equation}
Thus $A_j$ is fixed by the nonlocal current (up to a sign), and $A_j$ is positive wherever the Jost solutions remain
linearly independent \cite{Coz1970}.
In particular, the nonlocal-to-local transition is controlled by the behavior of the current: smooth, nonvanishing
$J_{N,j}$ leads to a well-behaved damping factor and a regular equivalent potential.

\subsection{Bilinear kernel $Q_{N,j}$ and the equivalent local potential}
Define the bilinear combination
\begin{equation}
Q_{N,j}(k;x,x')=-\frac{1}{2\ii k}\Big[\psi_{N,j,+}(k,x)\psi_{N,j,-}(k,x')
-\psi_{N,j,-}(k,x)\psi_{N,j,+}(k,x')\Big].
\label{eq:QNj}
\end{equation}
Then a local-equivalent potential can be written directly in terms of $J_{N,j}$, $Q_{N,j}$, and the kernel $U_j$ as
\begin{equation}
U^{\rm eq}_j(x;k)=
-\frac{1}{2}\frac{J_{N,j}''(k,x)}{J_{N,j}(k,x)}
+\frac{3}{4}\left(\frac{J_{N,j}'(k,x)}{J_{N,j}(k,x)}\right)^2
-\frac{1}{J_{N,j}(k,x)}\int_{0}^{\infty} \dd x'\,U_j(x,x')\,\partial_x Q_{N,j}(k;x,x').
\label{eq:Ueq}
\end{equation}
Equation~\eqref{eq:Ueq} yields an energy-dependent $V^{\rm eq}_j(x;\epsilon)=\hbar^2U^{\rm eq}_j(x;k)$ that is uniquely
fixed by the nonlocal kernel and the chosen nonlocal fundamental solutions \cite{Coz1970,Bhoi2022PhaseEquivalent}.
\par\noindent\textbf{Correspondence with Coz--Arnold--MacKellar (1970).}
With the identifications $r\mapsto x$, $V(r,r')/\hbar^2\mapsto U_j(x,x')$, and (for the half-line) $\ell=0$, our definitions reproduce the Coz mapping term by term: Eq.~\eqref{eq:JNdef} is the normalized current (their Eq.~(3.8)), Eq.~\eqref{eq:Aj2} is their damping identity $J_N=A^2$ (Eq.~(3.11)), and Eq.~\eqref{eq:Ueq} is the one-dimensional specialization of their equivalent local potential formula (Eq.~(3.17)). Consequently, whenever the Jost pair remains linearly independent (no current zeros), our component-wise local equivalents are exactly those of Ref.~\cite{Coz1970}.

From a practical viewpoint, \eqref{eq:Ueq} provides a direct diagnostic: if $J_{N,j}$ becomes small, the equivalent
potential becomes rapidly varying and the local picture loses predictive value.

\subsection{Spurious solutions and breakdown of the mapping}
The mapping fails at points where $J_{N,j}(k,x)=0$ (equivalently $A_j=0$), signalling loss of linear independence of
the incoming/outgoing solutions.
In this case the equivalent local potential can develop poles, and the corresponding states are the nonlocal
\emph{spurious solutions} discussed in the nonlocal Schr\"odinger literature \cite{Coz1970,FiedeldeySofianos1983}.
Operationally, spurious solutions correspond to energies at which the integral equation admits nontrivial solutions with
vanishing asymptotic flux, so that a local-equivalent interpretation in terms of standard scattering boundary
conditions becomes ill-defined.

\begin{prop}[Current criterion for validity of the local-equivalent map]
Assuming existence of Jost-type solutions $\psi_{N,j,\pm}$ on a domain $D\subseteq[0,\infty)$,
the local-equivalent map \eqref{eq:transform} is well-defined on $D$ if and only if
$J_{N,j}(k,x)\neq 0$ for all $x\in D$.
\end{prop}

\subsection{Reconstruction of an effective local generalized Dirac oscillator}
A local generalized Dirac oscillator is specified by a single function $f_{\rm eff}(x;\epsilon)$ producing partner
potentials
\begin{equation}
V^{\rm eff}_\pm(x;\epsilon)=f_{\rm eff}^2(x;\epsilon)\pm\hbar f_{\rm eff}'(x;\epsilon).
\label{eq:Veffpm}
\end{equation}
Given component-wise local equivalents $V^{\rm eq}_{1}(x;\epsilon)$ and $V^{\rm eq}_{2}(x;\epsilon)$ from
Eq.~\eqref{eq:Ueq}, define
\[
\Sigma(x;\epsilon):=V^{\rm eq}_{2}(x;\epsilon)+V^{\rm eq}_{1}(x;\epsilon),\qquad
\Delta(x;\epsilon):=V^{\rm eq}_{2}(x;\epsilon)-V^{\rm eq}_{1}(x;\epsilon).
\]
A necessary condition for the existence of a \emph{single} $f_{\rm eff}$ satisfying \eqref{eq:Veffpm} is
\begin{equation}
\Sigma(x;\epsilon)\ge 0\quad \text{and}\quad
\Delta(x;\epsilon)=2\hbar\, f_{\rm eff}'(x;\epsilon)\ \text{with}\ f_{\rm eff}^2(x;\epsilon)=\Sigma(x;\epsilon)/2.
\label{eq:compat-basic}
\end{equation}
Equivalently (away from zeros of $\Sigma$), the pair $(\Sigma,\Delta)$ must satisfy the differential compatibility
relation
\begin{equation}
\Delta(x;\epsilon)=
\pm\,\frac{\hbar\,\Sigma'(x;\epsilon)}{\sqrt{2\,\Sigma(x;\epsilon)}}.
\label{eq:compat-relation}
\end{equation}
If \eqref{eq:compat-relation} is violated, then the two component-wise local equivalents describe inequivalent local
oscillators and no single $f_{\rm eff}$ exists.
In applications, \eqref{eq:compat-relation} can therefore be used as a quantitative measure of how strongly the
nonlocal interaction breaks the simple SUSY pairing structure in a local representation.

\subsection{Finite-rank / separable example (Gaussian model and spurious criterion)}
A minimal and computationally tractable kernel model is a rank-one (separable) correction,
\begin{equation}
  f(x,x')= f_0(x)\,\delta(x-x')+\lambda\,u(x)u(x'),\qquad \lambda\in\mathbb{C},
\label{eq:separable}
\end{equation}
with a concrete choice such as a Gaussian profile on the half-line,
\begin{equation}
  u(x)=\exp\!\left(-\frac{x^2}{2a^2}\right)\chi_{[0,\infty)}(x),\qquad a>0.
\label{eq:gaussian-u}
\end{equation}
Finite-rank kernels are attractive because the integro-differential equation reduces to a system of ODEs coupled to a
finite number of moments (inner products) of the solution.
They also arise naturally as operator-theoretic approximations and in the study of singular (e.g.\ $\delta$-shell)
nonlocal interactions \cite{HeribanTusek2024}.

In the pure rank-one case $f_0\equiv 0$, the induced partner kernels $V_{1,2}$ remain finite rank and the component
problem reduces to an inhomogeneous ODE driven by $u$ and $u'$ together with a small linear system for the unknown
moments.
A spurious threshold is signaled by the vanishing of the corresponding finite-dimensional determinant; in that case
$J_{N,j}$ develops zeros and $U^{\rm eq}_j$ typically becomes singular.
The explicit reduction and the determinant criterion are collected in Appendix~A.4.

\subsection{Solvable benchmarks}
\paragraph{Example 1 (local Dirac oscillator).}
Take $f(x)=m\omega x$ (the Moshinsky--Szczepaniak Dirac oscillator) \cite{MoshinskySzczepaniak1989,Sadurni2011DMOReview}.
Then the partner potentials are
\[
V_-(x)=m^2\omega^2x^2-\hbar m\omega,\qquad
V_+(x)=m^2\omega^2x^2+\hbar m\omega,
\]
so the component equations reduce to shifted harmonic oscillators.
Using the standard oscillator spectrum $\hbar m\omega(2n+1)$ for
$-\hbar^2\,\mathrm{d}^2/\mathrm{d}x^2+m^2\omega^2x^2$, one obtains
\[
\epsilon_n^{(-)}=2\hbar m\omega\,n,\qquad
\epsilon_n^{(+)}=2\hbar m\omega\,(n+1),\qquad n=0,1,2,\dots
\]
with $\epsilon=(E^2-m^2c^4)/c^2$.
The Dirac coupling \eqref{eq:components-local} pairs the $n$-th level of $V_+$ with the $(n+1)$-th level of $V_-$,
so a convenient labeling yields
\begin{equation}
E_{n,\pm}=\pm\sqrt{m^2c^4+2\hbar\omega\,m c^2\,(n+1)},\qquad n=0,1,2,\dots,
\end{equation}
consistent with standard treatments (see also Ref.~\cite{Pacheco2003ThermalDO}).
At the level of eigenfunctions, one may choose $\psi_{2,n}(x)\propto\phi_n(x)$ and then
$\psi_{1,n}(x)=\tfrac{c}{E_{n,+}-mc^2}(p_x+\ii m\omega x)\psi_{2,n}(x)\propto\phi_{n+1}(x)$, making the inherited
ladder structure explicit.

\paragraph{Example 2 (translation-invariant nonlocal kernel).}
On the full line, choose a convolution kernel $f(x,x')=g(x-x')$ with $g\in L^1(\RR)$.
Then $(\partial_x+\partial_{x'})f(x,x')=0$, hence $[\hat F,p_x]=0$ and the partner operators coincide:
$A^\#A=AA^\#=p_x^2+\hat F^2$.
In momentum space, $\hat F$ is multiplication by $\tilde g(q)$, so plane waves $e^{\ii qx}$ diagonalize the problem with
\begin{equation}
\epsilon(q)=\hbar^2q^2+|\tilde g(q)|^2,\qquad
E_{\pm}(q)=\pm\sqrt{m^2c^4+c^2\epsilon(q)}.
\end{equation}
For these plane-wave fundamental solutions the normalized current is constant, $J_{N,j}\equiv 1$, so the damping is
trivial: $A_j\equiv 1$.
At a fixed energy (fixed $k^2=\epsilon/\hbar^2$), the local-equivalent potential becomes a constant chosen to reproduce
the same wave number $q$, namely $U^{\rm eq}=k^2-q^2$.
This illustrates that localization is most informative when the kernel breaks translation invariance (as in finite-range
optical-model nonlocalities), because otherwise the nonlocality primarily manifests as momentum (energy) dependence.

\paragraph{Example 3 (complex shift: pseudo-Hermitian but isospectral).}
Let $f(x)=m\omega(x-\ii a)$ with $a\in\RR$.
Then $f(x+\ii\hbar\theta)=f^*(x)$ holds for $\theta=a/\hbar$, and $H_{\rm GDO}$ is $\eta$-pseudo-Hermitian under
$\eta=e^{-\theta p_x}$ while remaining isospectral to the real Dirac oscillator up to a similarity transformation
\cite{Dutta2013,Mostafazadeh2010}.
More explicitly, $S=e^{-\theta p_x}$ acts as $(S\varphi)(x)=\varphi(x+\ii a)$ on sufficiently regular functions, so
$H_{\rm GDO}(a)=S\,H_{\rm GDO}(0)\,S^{-1}$ and the eigenfunctions follow by analytic continuation,
$\Psi_n^{(a)}(x)=\Psi_n^{(0)}(x+\ii a)$.
This local benchmark mirrors the nonlocal kernel criterion \eqref{eq:kernelcondition}: spectral reality is ensured by a
controlled complex translation at the level of the interaction profile.

% ==========================================================
\section{Conclusions and outlook}
\label{sec:concl}

We have developed a nonlocal extension of the generalized Dirac oscillator in $(1+1)$ dimensions by promoting the
interaction profile $f(x)$ to an integral operator with kernel $f(x,x')$.
At the operator level the resulting Dirac Hamiltonian preserves a natural factorization, which allows the full
first-order system to be reduced to two nonlocal Schr\"odinger-type partner equations for the spinor components.
We derived explicit coordinate-space formulas for the corresponding partner kernels, thereby making the induced
nonlocal supersymmetric structure completely transparent.

A central result is that the complex-translation metric $\eta=e^{-\theta p_x}$ leads to a simple \emph{kernel-level}
sufficient condition for $\eta$-pseudo-Hermiticity,
$f(x+\ii\hbar\theta,x'+\ii\hbar\theta)=f^*(x',x)$, which generalizes the familiar local complex-shift criterion for
pseudo-Hermitian Dirac oscillators.
This provides a practical way to generate nonlocal interactions with controlled spectral reality properties.

To connect the nonlocal dynamics with a local interpretation, we adapted the Coz--Arnold--MacKellar construction to each
component equation, yielding unique energy-dependent local-equivalent potentials and multiplicative damping (Perey)
factors whenever the nonlocal current remains nonzero.
The current criterion makes explicit the limitations of the mapping and ties its breakdown to current zeros and the
appearance of spurious solutions.

Several extensions merit further investigation.
These include: (i) systematic numerical studies for physically motivated kernels (e.g.\ Perey--Buck-type and modern
dispersive optical-model kernels) and a quantitative analysis of the resulting Perey suppression;
(ii) structural criteria and practical diagnostics for when the two component-wise local equivalents can be represented
by a single effective $f_{\rm eff}(x;\epsilon)$, thereby restoring a local GDO interpretation; and
(iii) generalizations to higher dimensions, to other choices of metric operators (including genuinely nonlocal metrics),
and to additional Lorentz structures beyond the pseudoscalar coupling considered here.

% ==========================================================
\appendix

\section{Extended proofs and detailed derivations}
For completeness we collect in this appendix several intermediate derivations used in the main text.
The material is organized in items A.1--A.4 to match the references in Sections~\ref{sec:nonlocal} and
\ref{sec:local-eq}.

\subsection{A.1 Proof of the partner-kernel formula \eqref{eq:V12kernels}}
\label{app:partner}
We derive the coordinate-space form of $\hat F^2$ and $[\hat F,p_x]$ used to obtain \eqref{eq:V12kernels}.

\paragraph{Step 1: $\hat F^2$.}
From \eqref{eq:Fkernel},
\[
(\hat F^2\phi)(x)=\hat F(\hat F\phi)(x)
=\int_{\RR}\dd y\,f(x,y)(\hat F\phi)(y)
=\int_{\RR}\dd y\int_{\RR}\dd x'\,f(x,y)f(y,x')\phi(x').
\]
Defining $(f\star f)(x,x'):=\int_{\RR}\dd y\,f(x,y)f(y,x')$ yields
$(\hat F^2\phi)(x)=\int_{\RR}\dd x'\,(f\star f)(x,x')\phi(x')$.

\paragraph{Step 2: the commutator $[\hat F,p_x]$.}
Compute
\[
(\hat F p_x\phi)(x)=\int_{\RR}\dd x'\,f(x,x')(-\ii\hbar)\partial_{x'}\phi(x').
\]
Assuming boundary terms vanish, integrate by parts in $x'$ to obtain
$(\hat F p_x\phi)(x)=\ii\hbar\int_{\RR}\dd x'\,(\partial_{x'}f(x,x'))\phi(x')$.
Next,
\[
(p_x\hat F\phi)(x)=(-\ii\hbar)\partial_x\int_{\RR}\dd x'\,f(x,x')\phi(x')
=-\ii\hbar\int_{\RR}\dd x'\,(\partial_x f(x,x'))\phi(x').
\]
Therefore
\[
([\hat F,p_x]\phi)(x)=\ii\hbar\int_{\RR}\dd x'\,(\partial_x+\partial_{x'})f(x,x')\phi(x'),
\]
which leads to \eqref{eq:V12kernels}.

\subsection{A.2 Proof of the kernel pseudo-Hermiticity condition \eqref{eq:kernelcondition}}
\label{app:pseudo}
Let $\eta=e^{-\theta p_x}$. Acting on a test function $\phi$,
\[
(\eta^{-1}\phi)(x)=\phi(x-\ii\hbar\theta),\qquad (\eta\phi)(x)=\phi(x+\ii\hbar\theta).
\]
Then
\[
(\eta\hat F\eta^{-1}\phi)(x)
=\eta\left(\int_{\RR}\dd x'\, f(\,\cdot\,,x')(\eta^{-1}\phi)(x')\right)(x)
=\int_{\RR}\dd x'\,f(x+\ii\hbar\theta,x')\,\phi(x'-\ii\hbar\theta).
\]
Assuming analyticity in the relevant strip and a legitimate contour shift, the change of variables
$x'\mapsto x'+\ii\hbar\theta$ yields \eqref{eq:etakernel}. Since
$(\hat F^\dagger\phi)(x)=\int\dd x'\,f^*(x',x)\phi(x')$, a sufficient condition for
$\hat F^\dagger=\eta\hat F\eta^{-1}$ is exactly \eqref{eq:kernelcondition}. Embedding $\hat F$ into
\eqref{eq:HNLGDO} then yields $H_{\rm NLGDO}^\dagger=\eta H_{\rm NLGDO}\eta^{-1}$.

\subsection{A.3 Proof of the current identity \eqref{eq:Aj2}}
Assume the multiplicative relation \eqref{eq:transform}.
Compute the Wronskian
\[
W_N:=\psi_{N,+}\partial_x\psi_{N,-}-\psi_{N,-}\partial_x\psi_{N,+}
=A^2\big(\psi_{L,+}\partial_x\psi_{L,-}-\psi_{L,-}\partial_x\psi_{L,+}\big),
\]
because the cross terms proportional to $AA'$ cancel.
Hence the normalized current $J_N=-(2\ii k)^{-1}W_N$ satisfies $J_N=A^2J_L$.
Since $J_L\equiv 1$ for the local equation with the chosen normalization, we obtain $J_N=A^2$.

\subsection{A.4 Separable-kernel reduction and an explicit spurious-solution trigger}
\label{app:separable}
To make the numerical program in Section~\ref{sec:local-eq} concrete, consider the pure rank-one case on the half-line:
\[
  f(x,x')=\lambda u(x)u(x'),\qquad x,x'\ge 0,
\]
with $u$ smooth and rapidly decaying.
Then
\[
  (f\star f)(x,x')=\int_0^{\infty}\dd y\,\lambda u(x)u(y)\,\lambda u(y)u(x')
  =\lambda^2 c_u\,u(x)u(x'),\qquad c_u:=\int_0^{\infty}\dd y\,u^2(y).
\]
Moreover,
\[
  (\partial_x+\partial_{x'})f(x,x')=\lambda\big(u'(x)u(x')+u(x)u'(x')\big).
\]
Therefore the partner kernels become
\[
V_{1,2}(x,x')=\lambda^2 c_u\,u(x)u(x')\mp\hbar\lambda\big(u'(x)u(x')+u(x)u'(x')\big),
\]
so the component equation \eqref{eq:nonlocal-schro} reads
\[
-\hbar^2\psi''(x)+\lambda^2 c_u\,u(x)S_0\mp\hbar\lambda\big(u'(x)S_0+u(x)S_1\big)=\epsilon\,\psi(x),
\]
where the two moments are
\[
S_0:=\int_0^{\infty}\dd x'\,u(x')\psi(x'),\qquad
S_1:=\int_0^{\infty}\dd x'\,u'(x')\psi(x').
\]
Thus the nonlocal problem reduces to a forced ODE with forcing in the span of $\{u,u'\}$ and two unknown constants
$(S_0,S_1)$.
Projecting onto $u$ and $u'$ yields a $2\times 2$ linear system
\[
\bm{M}_j(k)\begin{pmatrix}S_0\\ S_1\end{pmatrix}=\bm{b}_j(k),
\]
whose entries are explicit integrals involving the free Green function and $u,u'$ (details depend on the boundary
condition at $x=0$).
A \emph{spurious threshold} occurs when $\det\bm{M}_j(k)=0$: at such energies the Jost pair fails to be linearly
independent, $J_{N,j}(k,x)$ develops zeros, and the local-equivalent potential \eqref{eq:Ueq} becomes singular.

% ==========================================================

\end{document}